\documentstyle[aps,multicol]{revtex}

\input{tcilatex}

\begin{document}
\title{Microwave properties of superconducting $MgB_{2}$}
\author{N. Hakim, P.V.Parimi, C.Kusko, and S.Sridhar}
\address{Physics Department, Northeastern University, 360 Huntington Avenue,\\
Boston,MA 02115}
\author{P.C.Canfield, S.L.Bud'ko and D.K.Finnemore}
\address{Ames Laboratory and Department of Physics and Astronomy, Iowa State\\
University, Ames, IA 50011.}
\maketitle

\begin{abstract}
Measurements of the $10GHz$ microwave surface resistance of dense $MgB_{2}$
wire and pellet are reported. Significant improvements are observed in the
wire with reduction of porosity. The data lie substantially above the
theoretical estimates for a pure BCS s-wave superconductor . However the $%
R_{s}(20K)$ of the wire is an order of magnitude lower than that of
polycrystal $YBa_{2}Cu_{3}O_{6.95}$ and matches with single crystal $%
YBa_{2}Cu_{3}O_{6.95}.$The results show promise for the use of $MgB_{2}$ in
microwave applications.
\end{abstract}

\begin{multicols}{2}%

The discovery \cite{Akimitsu01} of superconductivity in $MgB_{2}$ with $%
T_{c}=40K$ has raised a flurry of interest. An important scientific issue is
whether $MgB_{2}$ is a conventional type II superconductor or falls in to a
different category, such as the high temperature superconducting cuprates
which are believed to have an exotic order parameter. A particularly
interesting technological possibility is the use of this material as a
replacement for $Nb$ in microwave applications such as superconducting
cavities and accelerators. A $40K$ s-wave superconductor would be of
enormous benefit for microwave applications since the microwave absorption
would be lower than the normal state by $10^{-5}-10^{-4}$ at $T\sim
T_{c}/2\sim 20K$, a temperature easily accessible with modern cryo coolers.

Microwave measurements provide accurate determination of applied and
fundamental parameters of superconductivity including $Z_{s}(T),$%
superconducting gap $\Delta (T)$, penetration depth $\lambda (T)$, pairing
symmetry, vortex dynamics and pinning force constant $k(T)$. Careful study
of superconductivity in $MgB_{2}$ is essential to determine the mechanism of
microwave absorption and thereby synthesizing better quality materials with
low microwave loss. In this report we present the first measurements of the
microwave surface impedance $Z_{s}=R_{s}-iX_{s}$ of $MgB_{2}$.

Two types of samples were studied. The first (P-1) is a polycrystalline
pellet ($\sim 1-2mm$ cubical dimensions) of moderate density which is
synthesized by reacting stoichiometric amounts of B and Mg at $950^{\circ }C$
for approximately two hours. The other W-1 is a high density $MgB_{2}$ wire
(outer diameter $140\mu m$) \cite{canfield01}. To synthesize $MgB_{2}$ wire
(W-1) powder Mg and a boron fiber of 100 $\mu m$ diameter, with central core
of Tungsten Boride (15 $\mu m)$, in a nominal ratio of $MgB_{2}$ are sealed
in to a $Ta$ tube. The sealed $Ta$ tube is then placed in $950^{\circ }C$ in
a box furnace for 2 hours and quenched to room temperature. The density of
the wire is at least 80\% of the theoretical density. Both samples P-1 and
W-1 have been extensively characterized by a variety of probes such as XRD,
resistivity, SEM and various other probes \cite{canfield01}. Very low dc
resistivity $\sim 0.4\mu \Omega -cm$ \cite{canfield01}at $T_{c}$ has been
observed in the wire which is lower than polycrystal samples by
approximately two orders of magnitude ($\sim 70\mu \Omega -cm$ \cite
{Akimitsu01}). The wire is more than $80\%$ dense, shows the full $\chi
=-1/4\pi \,$shielding in the superconducting state and has a sharp
transition of width $0.9K$ at $T_{c}=39.4K$.

The high precision microwave measurements were carried out in a $10GHz$ $Nb$
superconducting cavity with a variable temperature sample pedestal \cite
{zhaisri0}. This technique has been used extensively for characterizing the
microwave properties of small specimens of high $T_{c}$ and borocarbide
superconductors \cite{srikanth,tobi95}. The sample is placed in the maximum
of the microwave magnetic field of $TE_{011}$ mode and as it is warmed
slowly the surface impedance, $Z_{s}=R_{s}-iX_{s}$ is measured.

The superconducting microwave surface resistance $R_{s}(T)$ of P-1 and W-1
are shown in Fig. 1. The polycrystalline sample is observed to have higher $%
R_{s}$ when compared to that of wire. Both samples show approximately
similar temperature dependence of $R_{s}(T)$ but differ by a large scale
factor. The microwave resistivity $\rho _{\omega n}(T)$ (Fig. 2) in the
normal state is obtained from the surface resistance $R_{sn}(T)$ using $\rho
_{\omega n}(T)=2R_{sn}^{2}(T)/\mu _{0}\omega $. The data are in good
agreement with independent dc resistivity measurements of the wire. The
agreement between the dc and microwave resistivities confirms that the
classical skin-depth regime applies in the normal state.

Fig. 3 shows the scanning electron micrographs (SEM) of both P-1 and W-1
samples. The difference between $R_{s}$ of P-1 and W-1 can be attributed to
different densities and grain sizes of these two samples. As can be seen
from the figure the average grain size in W-1 is as large as $10$ $\mu m$
while P-1 has smaller grains of size $\sim 0.5-5\mu m.$ These differences in
grain size and porosity lead to larger absorption in P-1 in two ways.
Firstly, they result in coupled Josephson junctions with large effective
junction penetration depth, and secondly smaller grain size in P-1 increases
the effective surface area of and thereby $R_{s}$. In fact, correction to
surface area of P-1 by a factor of 10 does reduce $R_{s}$ reasonably close
to that of W-1. Fig.3(c) further clarifies the morphology of the $MgB_{2}$
sheath. Note that the Tungsten Boride core is not seen by the microwaves and
is shielded by $MgB_{2}$ because of the small penetration depth $\lambda
\sim 140nm$ \cite{Finnemore}, except for a small contribution at the wire
ends which may be responsible for some of the residual loss.

The comparisons to measurements at the same frequency for $Nb$, $Cu$ and
single crystal and polycrystal of $YBa_{2}Cu_{3}O_{6.95}$ \cite{srikanth}
are shown in Fig. 1. The wire sample starts off with a very low normal state 
$R_{s}$ slightly below $Cu$, and well below the normal state values for $Nb$%
. Very important to note is the comparison to $YBa_{2}Cu_{3}O_{6.95}$ where $%
R_{s}(T<T_{c})$ of $MgB_{2}$ wire is far below that of polycrystal $%
YBa_{2}Cu_{3}O_{6.95}$ and comparable to single crystal (Fig. 1) and films 
\cite{belk96,find96} of $YBa_{2}Cu_{3}O_{6.95}$ at $20K$. At present the
improvements in the superconducting state in $MgB_{2}$ are much smaller than
in the other superconductors. $Nb$ is well established to be an s-wave
superconductor and the microwave data are in excellent agreement with BCS
calculations of the microwave absorption for an s-wave superconductor with a
gap that is consistent with other measurements. On the other hand the origin
of the microwave absorption in $YBa_{2}Cu_{3}O_{6+x}$ is still debated, even
though a consensus has emerged that the order parameter is d-wave. In $%
YBa_{2}Cu_{3}O_{6+x}$ the microwave absorption is much larger than expected
even for a d-wave superconductor if one uses the quasiparticle scattering
time extrapolated from the normal state \cite{SridharHouston}.

Several reports have already claimed that $MgB_{2}$ is a conventional
phonon-mediated superconductor with an s-wave order parameter\cite{bud'ko}.
In conventional s-wave superconductors the thermodynamic and transport
coefficients decay exponentially and there is no quasiparticle excitations
at low energies. Tunneling measurements report the ratio $2\Delta
(0)/k_{B}T_{c}\sim 3$\cite{karapetrov01} in $MgB_{2}$. A conventional BCS
temperature dependence has been observed for the gap.

In order to investigate the mechanism of microwave absorption in $MgB_{2}$
the microwave data are compared with calculations for a BCS s-wave
superconductor. We use the relation between the impedance $Z_{s}=(-\mu
_{0}i\omega /\tilde{\sigma}_{s})^{1/2}$ and the Mattis-Bardeen complex
conductivity $\tilde{\sigma}_{s}=\sigma _{1}-i\sigma _{2}$. The normalized
conductivity can be calculated using $\sigma _{1}/\sigma _{n}=2/ 
\rlap{\protect\rule[1.1ex]{.325em}{.1ex}}h%
\omega \int_{\Delta }^{\infty }g(E)[f(E)-f(E+
\rlap{\protect\rule[1.1ex]{.325em}{.1ex}}h%
\omega )dE$ and $\sigma _{2}/\sigma _{n}=2/
\rlap{\protect\rule[1.1ex]{.325em}{.1ex}}h%
\omega \int_{\Delta -\omega }^{\Delta }h(E)[1-2f(E+ 
\rlap{\protect\rule[1.1ex]{.325em}{.1ex}}h%
\omega )dE$, where $g(E)$ and $h(E)$ are appropriate factors incorporating
the density of states and BCS coherence factors, and $f$ is the Fermi
function. The gap temperature dependence was taken to be the well-known gap
function for BCS. We have used experimental parameters $\omega =2\pi
10^{10}(\sec )^{-1}$ and $T_{c}=39.4K$. Using a weak coupling gap ratio $%
\Delta (0)/kT_{c}=1.76$, the calculated $R_{s}(T)$ is compared with the
measured data (Fig. 1). It is clear that the measured data are well above
the calculated values.

It is possible to arrive at a quantitative fit to the data by varying the
gap parameter $\Delta (0)/k_{B}T_{c}$. Including a residual temperature
independent contribution $R_{res}=150\mu \Omega $, the experimental data are
fit with $R_{s}(T)=R_{res}+R_{BCS-%
\mathop{\rm mod}%
}(T)$, where $R_{BCS-%
\mathop{\rm mod}%
}(T)\,$represents a modified BCS calculation in which the gap ratio $\Delta
(0)/kT_{c}\,$is variable in the calculations for $\sigma _{1}$ and $\sigma
_{2}$ mentioned above. Rather good fits are achieved with $\Delta
(0)/k_{B}T_{c}=0.18$. Such a low gap is clearly different from the tunneling
data \cite{karapetrov01}. The variable gap ratio in modified BCS calculation
is a convenient method of parametrization of the data, although it is
possible that there may be two energy scales in the material, with the
larger being observed in the tunneling experiments, while the smaller value
dominates the microwave absorption.

Interestingly the data for $MgB_{2}$ are similar to those in single crystals
of $RNi_{2}B_{2}C$ ($R=Y,Er,Ho,Dy,Tm)$ family of superconductors \cite
{tobi95}. The temperature dependence of microwave $R_{s}$ of $MgB_{2}$ and
its deviation from the BCS calculation for an s-wave superconductor is more
akin to the microwave $R_{s}$ of $RNi_{2}B_{2}C$ class superconductors which
also clearly do not follow BCS s-wave \cite{tobi95}. In both $MgB_{2}$ and $%
RNi_{2}B_{2}C$ the microwave $R_{s}(T)$ differs from the BCS calculation
with a broad temperature dependence near $T_{c}$ and high absolute values of 
$R_{s}$. The large microwave absorption observed in these materials was
previously attributed to pair breaking, which in the magnetic members of the
family could be possibly attributed to pair breaking due to magnetic ions%
\cite{tobi95}, although a possible additional mechanism due to strong
electronic correlations may be required in the non-magnetic member $%
YNi_{2}B_{2}C$. The absence of magnetic impurities in $MgB_{2}$ suggests a
non-magnetic mechanism. The intriguing similarity between $MgB_{2}$ and $%
RNi_{2}B_{2}C$ clearly deserves further attention.

In conclusion substantial improvements in $R_{s}$ of $MgB_{2}$ have been
achieved with improvement in sample density. A significant advantage of $%
MgB_{2}$ is the low normal state resistance due to the intermetallic nature
of the compound. The $R_{s}(20K)$ of W-1 is substantially lower than that of
polycrystal of $YBa_{2}Cu_{3}O_{6.95}$ and close to that of single crystal $%
YBa_{2}Cu_{3}O_{6.95}$. For a $10GHz$ $TE_{011}$ cavity constructed of $%
MgB_{2}$ with a geometric factor $\Gamma \sim 780$, and using $%
R_{s}(20K)=800\mu \Omega $, the resulting $Q(20K)=\Gamma /R_{s}>10^{6}$.
This observation is very promising, and further improvements in sample
quality would greatly enhance the prospects for microwave applications at
temperatures accessible using cost effective cryo coolers.

This work was supported at Northeastern by the Office of Naval Research.
Ames Laboratory is operated for the U.S.Department of Energy. The work at
Ames was supported by the Office of Basic Energy Sciences.

\narrowtext%

\begin{figure}
\caption{Microwave surface resistance of polycrystal pellet (P-1) and dense wire (W-1) of $MgB_2$.
For comparison, $R_s$ data are shown for  (a)  single crystal  $YBa_2 Cu_3 O_{6.95}$, 
(b) superconducting $Nb$, (c)   polycrystal $YBa_2 Cu_3 O_{6.95}$, and (d) $Cu$. 
(e) Solid line represents $R_s$ for a BCS s-wave superconductor with gap ratio $\Delta (0)/kT_c = 1.76$.(f) LIght gray line
represents a modified BCS calculation.}
\end{figure}
%

\begin{figure}
\caption{Microwave resistivity $\rho_{\omega n}(T)$ of pellet P-1 and wire W-1of $MgB_2$. }
\end{figure}
%

\begin{figure}
\caption{SEM micrographs of pellet  P-1 (a) and wire  W-1 (b $\&$ c).
 P-1 is highly porous and results in weakly connected small grains.
 Large well aligned grains of average size 10 $\mu m$ can be seen in W-1.
 (c) Wire W-1 cut through the midsection. Note that the Tungsten Boride core is shielded from
the microwave fields by the $MgB_2$ sheath.}
\end{figure}
%

\end{multicols}%

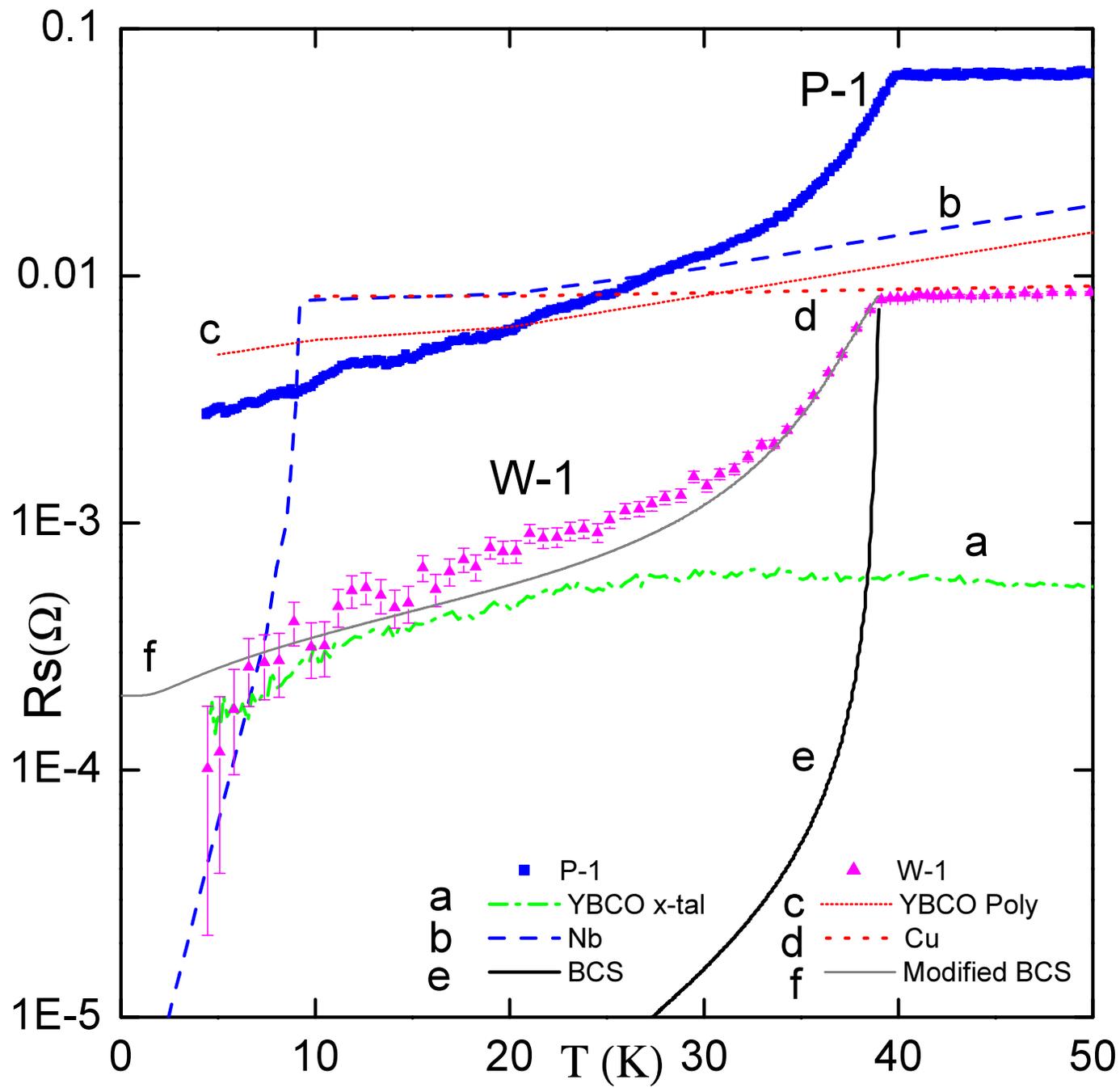
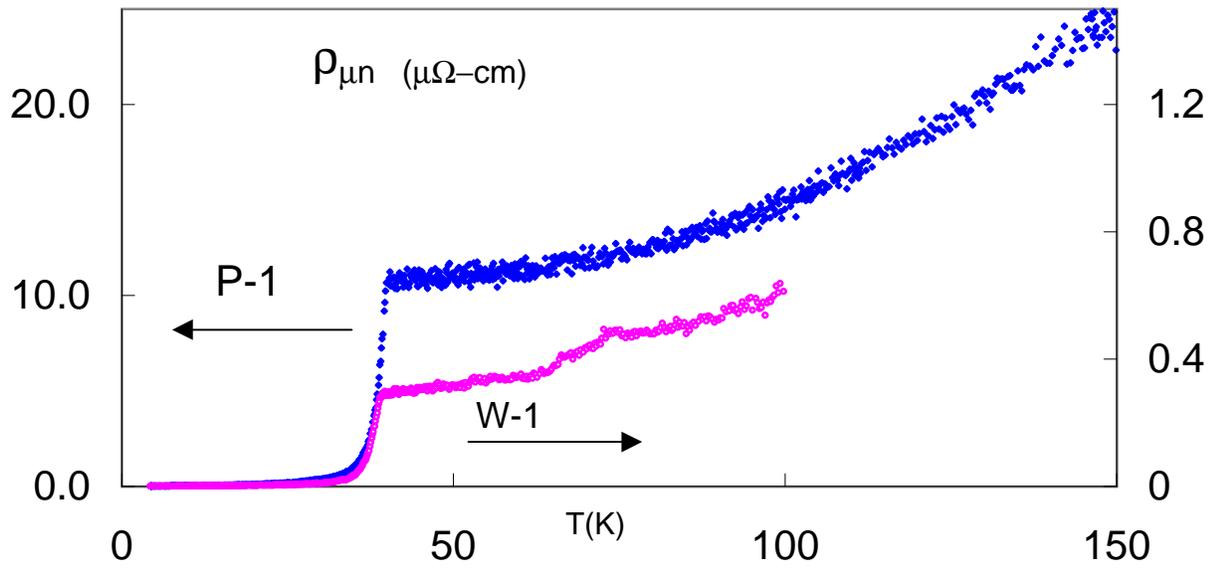
\end{document}